\documentstyle[12pt,aps]{revtex}

\input psfig
\begin{document}
\baselineskip=17pt

\draft
\title{High Energy Neutrinos from Astrophysical Sources: An Upper Bound}
\author{Eli Waxman and John Bahcall}
\address{Institute for Advanced Study, Princeton, NJ 08540; E-mail:
waxman@sns.ias.edu, jnb@sns.ias.edu}
\date{\today}
\maketitle
\begin{abstract}
\tightenlines

We show that cosmic-ray observations set a model-independent upper
bound of $E_\nu^2\Phi_\nu<2\times10^{-8}{\rm GeV/cm}^2{\rm s\,sr}$ 
to the intensity 
of high-energy neutrinos produced by photo-meson (or p-p) interactions
in sources 
of size not much larger than the proton photo-meson (or p-p)
mean-free-path.
This bound applies, in particular,
 to neutrino production by either AGN jets or  GRBs. 
The upper limit is two orders
of magnitude below the intensity predicted in some popular AGN jet models
and therefore contradicts the 
theory that the cosmic gamma-ray background is due
to photo-pion interactions in AGN jets. The upper bound 
is consistent with our predictions from GRB models. 
The predicted intensity 
from GRBs is $E^2 dN/dE\sim0.3\times10^{-8}{\rm GeV/cm}^2{\rm s\,sr}$ for 
$10^{14}{\rm eV}<E<10^{16}{\rm eV}$; we also derive the expected intensity at 
higher energy. 
\end{abstract}


\section{Introduction}

Large area, high-energy neutrino telescopes are being constructed to detect 
cosmologically distant neutrino sources (see \cite{GHS} for a review). 
The main motivation of a search for cosmological high-energy neutrino sources
derives from the fact that the cosmic-ray energy spectrum extends to
$>10^{20}$eV and is most likely dominated above $\sim3\times10^{18}$eV by
an extra-Galactic source of protons \cite{WatsonFA}. 
High-energy neutrino production is likely 
to be associated with the production of
high-energy protons, through the decay of charged pions produced by 
photo-meson interaction of the high-energy protons with the radiation field
of the source.
Gamma-ray bursts (GRB) \cite{WnB} and active galactic nuclei (AGN) jets 
\cite{AGNnu} have been 
suggested as possible sources of high-energy neutrinos that are
associated with high-energy cosmic-rays. 
The predicted neutrino fluxes may be detectable with high-energy 
neutrino telescopes of effective area $\sim1{\rm\ km}^2$.

We show here that high-energy cosmic-ray observations
set a model-independent upper bound to the expected 
high-energy neutrino fluxes
and are in conflict with the theory that the 
cosmic gamma-ray
background is due to photo-pion interactions in AGN jets.  The
upper bound is stated in Eq. (\ref{Jnu}) and is illustrated in Fig.~1.
The demonstration that the AGN jet models for the gamma-ray 
background are in conflict with the cosmic ray data is given in Section III.
We also give a more detailed prediction 
than we gave in ref. \cite{WnB} for the expected GRB neutrino spectrum
and discuss the compatibility of the detailed results
with the general secondary-particle cooling 
constraints derived by Rachen \& M\'esz\'aros \cite{RM98}.

It has been suggested that neutrinos may be produced
in the cores of AGNs (rather than in the jets) 
by photo-meson interaction of protons accelerated to
high energy in the AGN core; in this scenario neutrinos are produced very 
close to the central black-hole \cite{Stecker}. In this model, the proton 
photo-meson optical depth is very high, $\tau_{p\gamma}\sim100$, and 
high-energy nucleons do not escape the source. 
By construction, there can be no observational evidence except neutrinos
for, or against,
the hypothesized luminous AGN accelerator of high-energy protons.
The hypothesized black-hole accelerators are ``neutrino-only'' factories.
Therefore, cosmic-ray observations
can not set a limit to neutrino emission in this model. On the other-hand,
this model can not  
explain, and is therefore not supported by, the existence of the 
extra-Galactic high-energy cosmic-ray flux.

This paper is organized as follows. In section II we derive the general 
upper bound to neutrino fluxes from $p-\gamma$ interactions for sources 
optically thin to $p-\gamma$ reactions. 
We compare in this section the upper limit to the predictions from
different models for neutrino sources. We also show that the upper
bound cannot be avoided by cosmological 
evolutionary effects (\S IIc) or by invoking magnetic fields (\S III).
In section IV we discuss the 
implications of the upper bound for  AGN jet models
of neutrino production. In section V we discuss the implications 
for  the GRB model of high energy neutrino production, 
derive a more detailed prediction (compared to our
prediction in ref. \cite{WnB}) for the expected GRB neutrino spectrum, and
compare our results with those of other authors. 
We discuss in Section VI our  main conclusions.

\section{Upper bound to the neutrino flux}
\label{sec:upperbound}

We first derive in section IIA the upper bound to the high-energy neutrino
flux from the sources at redshift $z<1$ that produce the observed cosmic-rays
at energies greater than $10^{18}$~eV. We compare in section IIB the upper 
bound with the predictions of current models. In section IIC we discuss the 
modification of the upper bound by unobserved sources of cosmic rays at
larger redshift. 

\subsection{Derivation of the upper bound}

Cosmic-ray observations above $10^{17}$eV indicate that an extra-Galactic 
source of protons dominates the cosmic-ray flux
above $\sim3\times10^{18}{\rm eV}$ \cite{WatsonFA}, while
the flux at lower energies is dominated by heavy ions, most likely of 
Galactic origin. 
The observed energy spectrum of the extra-Galactic component is 
consistent with that expected for a cosmological
distribution of sources of protons, with injection spectrum
$dN_{CR}/dE_{CR}\propto E_{CR}^{-2}$, as typically
expected for Fermi acceleration \cite{CRflux}. 
The energy production rate of protons in the energy range
$10^{19}$eV to $10^{21}$eV is 
$\dot\varepsilon_{CR}^{[10^{19},10^{21}]}\sim5\times10^{44}
{\rm erg\ Mpc}^{-3}
{\rm yr}^{-1}$ \cite{CRflux}, if the observed
flux of ultra-high-energy cosmic-rays is produced by 
sources that are cosmologically distributed. 
The energy-dependent generation rate of cosmic-rays
is therefore given by 
\begin{equation}
E_{CR}^2{d\dot N_{CR}\over dE_{CR}}={\dot\varepsilon_{CR}^
{[10^{19},10^{21}]}\over
\ln(10^{21}/10^{19})}\approx
10^{44}{\rm erg\ Mpc}^{-3}{\rm yr}^{-1}.
\label{ECR}
\end{equation} 
If the high-energy
protons produced by the extra-galactic sources lose a fraction
$\epsilon<1$ of their
energy through photo-meson production of pions before escaping the source, 
the resulting present-day energy density of muon neutrinos is 
$E_\nu^2 dN_\nu/ dE_\nu
\approx0.25\epsilon t_H E_{CR}^2d\dot N_{CR}/dE_{CR}$, 
where $t_H\approx10^{10}{\rm yr}$ is the Hubble time.
For energy independent $\epsilon$ 
the neutrino spectrum follows the proton generation spectrum, since
the fraction of the proton energy carried by a neutrino produced through a
photo-meson interaction, $E_\nu\approx0.05E_p$, 
is independent of the proton energy.
The $0.25$ factor arises because neutral pions, 
which do not produce neutrinos, are produced with roughly equal
probability with charged pions, and because in the decay 
$\pi^+\rightarrow\mu^++\nu_\mu
\rightarrow e^++\nu_e+\overline\nu_\mu+\nu_\mu$
muon neutrinos carry approximately half
the charged pion energy. Defining $I_{\rm max}$ as the muon neutrino intensity 
($\nu_\mu$ and $\bar\nu_\mu$ combined) obtained for $\epsilon=1$, 
\begin{equation}
I_{\rm max}\approx0.25\xi_Zt_H{c\over4\pi}E_{CR}^2{d\dot N_{CR}\over dE_{CR}}
\approx1.5\times10^{-8}\xi_Z{\rm GeV\,cm}^{-2}{\rm s}^{-1}
{\rm sr}^{-1}\,,
\label{Fmax}
\end{equation}
the expected neutrino intensities are
\begin{equation}
E_\nu^2\Phi_{\nu_\mu}\equiv{c\over4\pi} E_\nu^2{dN_{\nu_\mu}\over dE_\nu}
={1\over2}\epsilon I_{\rm max},\quad 
\Phi_{\nu_e}\approx\Phi_{\bar\nu_\mu}\approx\Phi_{\nu_\mu}.
\label{Jnu}
\end{equation}
The quantity $\xi_Z$ in Eq. (\ref{Fmax}) is of order unity and has been 
introduce here to describe the possible contribution
of so far unobserved high redshift
sources of high-energy cosmic rays and to include the effect of the redshift
in neutrino energy. We estimate $\xi_Z$ in section IIC.

\subsection{Upper bound versus current models}

Figure 1 compares the neutrino intensity predictions of GRB and AGN jet
models with the intensity given by Eq. (\ref{Fmax}).
The AGN core model predictions are shown for completeness. The intensities 
predicted by both AGN jet and core models exceed $I_{\rm max}$ by
typically two orders of magnitude.

\begin{figure}[!ht]
\tightenlines
\centerline{\psfig{figure=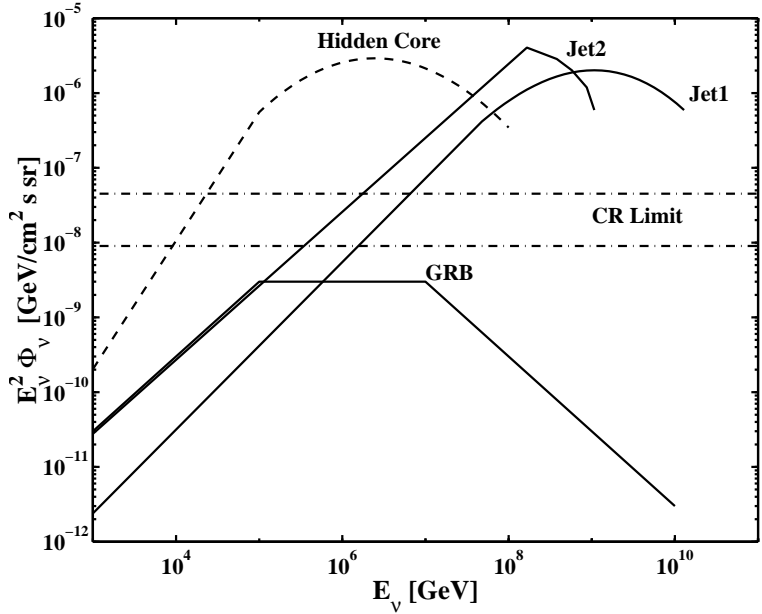,width=4.in}}
\caption[]{
Comparison of muon neutrino intensities 
($\nu_\mu$ and $\bar\nu_\mu$ combined) predicted by different models with the 
upper bound implied by cosmic ray observations. The 
dash-dot lines give the upper bound Eq. (\ref{Fmax}) corrected
for neutrino energy loss due to redshift and for possible redshift evolution
of the cosmic-ray generation rate. The lower line is obtained assuming no
evolution, and the upper line assuming rapid evolution similar to the 
evolution of the QSO luminosity density. The AGN jet model predictions
are taken from ref. \cite{AGNnu} (labeled 'Jet1' and 'Jet2'). 
The GRB intensity is based on the estimate presented in this paper, following
\cite{WnB}. The AGN hidden-core 
conjecture, which produces only neutrinos and 
to which the upper bound does not apply, 
is taken from \cite{Stecker}.
}
\label{fig1}
\end{figure}

The intensity $I_{\rm max}$ is an upper bound to the intensity of high-energy
neutrinos produced by photo-meson interaction in sources
of size not much larger than the proton photo-meson mean-free-path. Higher
neutrino intensities from such sources would imply proton fluxes higher than 
observed in cosmic-ray detectors. Clearly, higher neutrino intensities may
be produced by sources where the
proton photo-meson ``optical depth'' is much
higher than unity, in which case only the neutrinos escape the source. However,
the existence of such sources cannot be motivated by the observed high-energy
cosmic-ray flux or by any observed electromagnetic radiation. We therefore
refer in Fig. 1 to models with $\tau_{\gamma p}\gg1$ as ``hidden core'' 
models.

\subsection{Evolution and redshift losses}

In the derivation of Eq. (\ref{Fmax}) we have neglected the redshift energy
loss of neutrinos produced at cosmic time $t<t_H$, 
and implicitly assumed that the cosmic-ray 
generation rate per unit (comoving) volume is independent of cosmic
time. The generation rate may have been higher at earlier times, i.e.
at high redshift. Cosmic rays above $10^{18}$~eV must originate from sources
at $z<1$. Energy loss due to redshift and pair production in interaction with
the microwave background implies that in order to be observed with energy
$E>10^{18}$eV, a proton should have been produced at $z=1$ with energy
exceeding the threshold for photo-meson production in interaction with
the microwave background at that redshift. Photo-meson energy loss of protons 
produced above the threshold would reduce the proton energy to the threshold
value in a short time, so that their observed energy (i.e. at $z=0$) would
be $\sim10^{18}$~eV. Thus, the cosmic ray energy generation rate given in
Eq. (\ref{ECR}) is the present (i.e low redshift, $z<1$) generation rate.
An increase in the cosmic-ray energy
generation rate per unit (comoving) volume above the value of Eq. 
(\ref{ECR}) at large redshift, $z>1$, is consistent with observations, since
it would not affect the observed flux above $\sim10^{18}$~eV, and the 
contribution from $z>1$ sources to the observed flux below $\sim10^{18}$~eV
may be hard to detect due to the ``background'' of Galactic sources of
heavy ion cosmic-rays which are most likely dominating the flux at this energy
\cite{WatsonFA,aniso}. 

Let us consider the possible modification of Eq. (\ref{Fmax}) due to evolution
and redshift losses. A neutrino with observed energy $E$ must be
produced at redshift $z$ with energy $(1+z)E$. Thus, 
the present number density of neutrinos above energy $E$ is given by
\begin{equation}
n_\nu(>E)=\int_0^{z_{\rm max}} dz{dt\over dz}\dot n_\nu[>(1+z)E,z]=
          \dot n_0(>E)\int_0^{z_{\rm max}} dz{dt\over dz}(1+z)^{-1}f(z).
\label{zint}
\end{equation}
Here we have used the fact that $\dot n_\nu(>E)\propto E^{-1}$ and
denoted the ratio of (comoving) neutrino production rate at redshift
$z$ to the present rate, $\dot n_0$, by $f(z)$. Comparing Eqs. (\ref{Fmax}) and
(\ref{zint}), and noting that $t_H\equiv\int_0^{\infty} dz (dt/dz)$, we
find that the intensity $I_{\rm max}$ 
of Eq. (\ref{Fmax}) should be multiplied by a correction factor
\begin{equation}
\xi_Z={\int_0^{z_{\rm max}} dz g(z)(1+z)^{-7/2}f(z)
     \over \int_0^\infty dz g(z)(1+z)^{-5/2}}.
\label{corr}
\end{equation}
Here, $g(z)\equiv-H_0(1+z)^{5/2}(dt/dz)$ is a weak function
of redshift and cosmology; $g(z)\equiv1$ 
for a flat universe with zero cosmological constant. Let us assume that
the neutrino energy generation rate evolves rapidly with redshift, following
the luminosity density evolution of QSOs \cite{QSO}, which may be described
as $f(z)=(1+z)^{\alpha}$ with $\alpha\approx3$ \cite{QSOl} at low redshift, 
$z<1.9$, $f(z)={\rm Const.}$ for $1.9<z<2.7$, and an exponential decay at 
$z>2.7$ \cite{QSOh}. Using this functional form of $f(z)$, which is 
also similar \cite{QSO} to that 
describing the evolution of star formation rate \cite{SFR}, we find that
$\xi_Z\approx3$ (with weak dependence on cosmology). For no evolution,
$f(z)={\rm Const.}$, we have $\xi_Z\approx0.6<1$ 
(with only a weak dependence on cosmology) due to a redshift energy loss of
neutrinos.

\section{Can the upper bound be avoided by invoking magnetic
fields?}
\label{magnetic}

One may try to invent arguments in order to avoid
the upper bound on the neutrino flux by hypothesizing strong magnetic fields, 
which would affect the propagation of cosmic-ray protons,
either in the neutrino source or in the inter-galactic medium 
between the source and the Earth. We show in this section that magnetic
fields in the source can not affect the upper bound, and that observational 
constraints on inter-galactic magnetic fields imply that proposed 
field scenarios also do not affect the upper bound.

\subsection{Magnetic fields in the neutrino source}

One might try to argue that the upper bound $I_{\rm max}$ 
can be avoided even for sources with small photo-meson optical
depth, if protons are prevented from escaping the source by magnetic
confinement. However, a photo-meson interaction producing a charged pion
also converts the proton to a neutron, which is not magnetically confined
and will escape a source with small photo-meson optical depth before
decaying to a proton.
A neutron of high energy $E$ propagates a distance $100(E/10^{19}{\rm eV})
{\rm kpc}$ prior to its decay. 
Thus, magnetic fields within the neutrino source cannot be used to
evade the upper bound given in Eq. (\ref{Fmax}).

\subsection{Uniformly distributed inter-galactic magnetic fields}

The existence of uniformly distributed inter-galactic magnetic fields
may limit the propagation distance of protons and therefore prevent their
arrival at Earth from distant sources. However, 
imposing a limit on  the propagation distance of
cosmic-rays will not affect any of the arguments presented in
this paper. Compare, e.g., the case where protons propagate on straight lines 
to the case where cosmic rays 
are confined by magnetic fields to their place of 
origin. In the former case, the present proton flux is obtained
by integrating the contribution of 
distant sources over redshift, while in the latter case it is obtained by 
integrating the contribution from local sources over cosmic time. 
For a homogeneous universe both procedures  yield the same 
result  [e.g. the integration in Eq. (\ref{zint})
may be interpreted as integration over redshift or over time]. 
Thus, limiting the cosmic-ray propagation 
distance may affect the upper bound 
$I_{\rm max}$  only if propagation is limited
to a distance $d$ over which the cosmic ray
production rate, averaged over a Hubble time, is inhomogeneous. In this case,
if the cosmic-ray production rate in our local region of size $d$ 
is lower than the universe average, the observed cosmic-ray flux would be 
lower than average and the neutrino flux, which is homogeneous throughout the
universe, may
exceed a bound based on the local cosmic-ray flux. However, the magnetic
fields required to make such a scenario viable, i.e. to confine cosmic rays 
to small enough $d$ to significantly affect $I_{\rm max}$, 
are large and are inconsistent with observations.

Consider a proton of energy $E$ propagating through an intergalactic 
magnetic field of
strength $B$ and correlation length $\lambda$. Propagating a distance
$\lambda$ the proton is deflected by an angle $\sim\lambda/R_L$, where
$R_L=E/eB$ is the Larmor radius. For the parameters of interest (see below)
the deflection angle is small, and propagating a distance $l$ the proton
is deflected by an angle $(l/\lambda)^{1/2}\lambda/R_L$. Thus, we
may define an effective mean free path, the propagation distance over
which large deflection occurs, by $(l\lambda)^{1/2}/R_L=1$ and a 
diffusion coefficient for proton propagation, $D=lc/3$. For a propagation time
$t$, protons are confined to a region of size $d\sim(Dt)^{1/2}=
(ct/3\lambda)^{1/2}R_L$. For $t=t_H\approx10^{10}$~yr, we have
$d\sim1(E/3\times10^{19}{\rm\,eV})(B_{\rm nG}
\lambda_{\rm Mpc}^{1/2})^{-1}$~Gpc.
The propagation distance is determined by the product $B\lambda^{1/2}$.
The upper limit on the intergalactic magnetic field implied by QSO Faraday 
rotation measurements, $B\lambda^{1/2}<1
{\rm\, nG\,Mpc}^{1/2}$ \cite{field1,field2}, implies
$d>1(E/3\times10^{19}{\rm\,eV})$~Gpc. 
We conclude that 
the existence of a uniformly distributed inter-galactic magnetic field would
have no effect on $I_{\rm max}$. 

\subsection{Magnetic fields in large scale structures}

The discussion in the previous subsections shows that magnetic field in the 
source or a uniformly distributed inter-galactic field would not 
affect the upper bound $I_{\rm max}$. 
Could magnetic fields associated with large scale galaxy structures, 
i.e. clusters, filaments and sheets, affect the upper bound?

Let us first consider galaxy clusters, where inter-galactic fields
had been detected. The analysis of rotation measures of radio sources
lying in the background of rich clusters implies the existence of strong
fields between galaxies, $B\sim1\mu$G with $\lambda\sim10$~kpc, 
in the central $0.5$~Mpc, cluster region \cite{Kim91}. 
However, confinement of
high energy protons in the cores of rich clusters would have little effect
on our results, unless most of the high energy neutrino sources reside
in the centers of rich clusters, which is not the case for either the 
hypothesized AGN or GRB neutrino sources. Moreover, 
high-energy protons can not be confined even in the central regions of
a rich cluster, since for $B\sim1\mu$G and $\lambda\sim10$~kpc
we have $d\sim10(E/3\times10^{19}{\rm\,eV})$~Mpc over a Hubble time. 

We next turn to large-scale filaments/sheets.
Kulsrud {\it et al.} \cite{Kulsrud} has recently suggested
that magnetic fields could be amplified by turbulence associated with
the formation of large scale filaments and sheets to near equipartition with 
turbulent kinetic energy. For characteristic turbulent velocities of 
$v_t\sim300{\rm\ km/s}$ on $\sim1$~Mpc scale, and characteristic filament/sheet
densities of $n\sim10^{-6}{\rm\ cm}^{-3}$, this scenario predicts magnetic
fields $B\sim0.1(n/10^{-6}{\rm\ cm}^{-3})^{1/2}(v_t/300{\rm\ km/s})
\ \mu{\rm G}$ in the high density large scale filaments/sheets, with coherence
length $L\sim1$~Mpc, comparable to the filament (sheet) diameter (thickness).
It is not clear whether the suggested scenario for increase in magnetic field
strength and coherence length to equipartition with the largest turbulent
eddies can be realized. 
Furthermore, even for a turbulent velocity of order several
hundreds km per sec, the turn-around time of a $\sim 1$~Mpc eddy is longer 
than the Hubble time, and it is therefore not clear whether equipartition with
the largest scale is achievable. Nevertheless, we consider this scenario
here since it is consistent with the upper limit, $B<1\ \mu{\rm G}$, 
implied for a field coherent over $\sim1$~Mpc
inside high density large scale structures by QSO rotation measures
\cite{Ryu}. Although the Larmor radius of a $\sim10^{19}$~eV proton
is smaller than $L\sim1$~Mpc, confinement of particles would require 
a special
field configuration. Even if such a configuration is produced by the random
turbulent motions, which seems unlikely, variation of the field over scale
$L$ gives rise to particle drift velocity, $v_d\sim R_L c/L$, and 
therefore to the escape of particles on time scale $t_e\sim L^2/cR_L=
10^7(L/1{\rm\ Mpc})^2(B/0.1\mu{\rm G})(E/3\times10^{19}{\rm eV})$~yr.
Since $t_e\ll t_H$, the hypothesized large-scale structure magnetic fields
can not affect the bound on neutrino flux.

Finally, we note that several authors have recently considered cosmic-ray
proton propagation in an hypothesized large scale  
magnetic field, associated with our local super-cluster, of $0.1\mu{\rm G}$ 
strength and 10 Mpc coherence length corresponding to a hypothetical local 
turbulent eddy of comparable size \cite{Biermann,Tanco}. For these parameters
as well we have $t_e\ll t_H$, and therefore even this
field structure would not affect the neutrino bound. Nevertheless, two points
should be made. First, if cosmic-rays are confined to our local super-cluster,
one would expect the local cosmic-ray 
flux to be higher than average (since the 
production rate averaged over Hubble time should be higher than average in
over-dense large scale regions), implying 
that the upper bound on the neutrino intensity is lower than $I_{\rm max}$ derived
here. Second, it is hard to understand how the hypothesized magnetic field 
structure could have been formed. The overdensity in the local super-cluster
is not large and an equipartition magnetic field of strength $0.1\mu{\rm G}$
therefore corresponds to a turbulent velocity $v_t\sim10^3{\rm\ km/s}$. A
turbulent eddy of this velocity coherent over tens of Mpc is inconsistent
with local peculiar velocity measurements (e.g. \cite{Strauss} for 
a recent review). Moreover, the corresponding eddy turn around time
is larger than the Hubble time.

\section{AGN jet models}

We consider in this section some popular models for neutrino production in
which high-energy neutrinos are produced in the jets of active galactic
nuclei \cite{AGNnu}. 
In these models, the flux of high-energy neutrinos received at Earth is 
produced by ``blazars'', AGN
jets nearly aligned with our line of sight.
Since the predicted neutrino 
intensities for these models exceed by typically two
orders of magnitude the upper bound, Eq. (\ref{Fmax}), based on observed
cosmic ray fluxes, it is important to verify that the models satisfy
the assumption on which Eq. (\ref{Fmax}) is based, i.e. optical depth $<1$
to $p-\gamma$ interaction. 

The neutrino spectrum and flux are derived in AGN jet models on the basis of
the following key considerations. 
It is assumed that protons are
Fermi accelerated in the jet to high energy, with energy spectrum 
$dN_p/dE_p\propto E_p^{-2}$. For a photon spectrum $dN_\gamma/dE_\gamma
\propto E_\gamma^{-2}$, as typically observed, the number of photons with
energy above the threshold for pion production is proportional to the
proton energy $E_p$ (the threshold energy is inversely proportional
to $E_p$). This implies that the proton photo-meson optical depth is
proportional to $E_p$, and therefore, assuming that the optical depth
is small, that the resulting neutrino
spectrum is flatter than the proton spectrum, namely
$dN_\nu/dE_\nu\propto E_\nu^{-1}$, as shown in Fig. 1. The spectrum
extends to a neutrino energy which is $\approx5\%$ of the maximum accelerated
proton energy, which is typically $10^{19}$eV in the models discussed.

The production of charged pions is accompanied by the production of
neutral pions, whose decay leads to the emission of high-energy gamma-
rays. It has been claimed \cite{pionAGN} 
that the observed blazar emission extending
to $\sim10$~TeV \cite{TeV} supports the
hypothesis that the high-energy emission is due to neutral pion decay
rather then to inverse Compton scattering by electrons.
Thus, the normalization of the neutrino flux is determined by the assumption 
that neutral pion decay is the source of high-energy
photon emission and that this emission from AGN jets produces the
observed  diffuse $\gamma$-ray background, $\Phi_\gamma(>100{\rm MeV})=10^{-8}
{\rm erg/cm}^2{\rm s\,sr}$ \cite{EGRET}. Under these assumptions
the total neutrino energy flux is similar to the $\gamma$-ray background
flux (see Fig. 1).

In the AGN jet models discussed above, the proton photo-meson 
optical depth $\tau_{p\gamma}$ at $E_p\le10^{19}$eV is smaller than unity. 
This is evident from the neutrino energy spectrum shown in Fig. 1, which is
flatter than the assumed proton spectrum at $E_p\le10^{19}$, as explained
above. In fact, it is easy to see that these models are constrained to
have $\tau_{p\gamma}\le10^{-3}$ at $E_p\sim10^{19}$eV. 
The threshold energy of photons for pair-production
in interaction with a 1~TeV photon is similar to the photon energy
required for resonant meson production in interaction with a proton of energy
$E_p=0.2{\rm GeV}^2/(0.5{\rm MeV})^2\times1\,{\rm TeV}=10^{18}$~eV.
Emission of $\sim1$~TeV 
photons from blazars is now well established \cite{eTeV},
and there is evidence that the high-energy photon spectrum extends as a 
power-law at least to $\sim10$~TeV \cite{TeV}. 
This is the main argument used \cite{pionAGN} in 
support of the hypothesis that high-energy emission from blazars
is due to pion decay rather than inverse Compton scattering.
The observed high-energy emission implies that the pair-production optical 
depth for $\sim1$~TeV photons is small, and that
$\tau_{p\gamma}\le10^{-4}(E_p/10^{18}{\rm eV})$, since the cross
section for pair production is $\sim10^4$ times larger than the cross section 
for photo-meson production. 
This result guarantees that the upper bound $I_{\rm max}$ on the neutrino
intensity is valid for AGN jet models.

\section{Gamma-ray bursts}

In the GRB fireball model for high-energy neutrinos, the cosmic ray 
observations are naturally taken into account and the upper limit on 
high energy neutrino flux  is automatically satisfied. In fact, it was the 
similarity between the energy density in cosmic ray sources implied by the 
cosmic ray  flux observations and the GRB energy density in high energy 
protons that led to the initial suggestion that GRBs are the the source of 
high energy protons. Just as for AGN jets, the GRB fireballs are optically 
think to $\gamma$-
p interactions that produce pions but--unlike the AGN jet models--the GRB 
model predicts a neutrino flux that satisfies the cosmic-ray upper bound 
discussed in Section II.

\subsection{Neutrinos at energies $\sim10^{14}$~eV}

In the GRB fireball model \cite{fireball}, which
 has recently gained
support from GRB afterglow observations \cite{AG}, the observed gamma rays 
are produced by synchrotron emission of high-energy electrons accelerated in
internal shocks of an expanding relativistic wind, with characteristic
Lorentz factor $\Gamma\sim300$ \cite{lorentz}. 
In this scenario, observed gamma-ray flux 
variability on time scale $\Delta t$ is produced by internal collisions at 
radius $r_d\approx\Gamma^2c\Delta t$ that arise from
variability of the underlying source on the same time scale \cite{internal}.
In the region where electrons are accelerated, protons are also expected to be
shock accelerated, and their photo-meson interaction with observed burst
photons will produce a burst of high-energy neutrinos accompanying
the GRB \cite{WnB}. If GRBs are the sources of
ultra-high-energy cosmic-rays \cite{GRB1,GRB2}, 
then the expected GRB neutrino intensity is \cite{WnB}
\begin{eqnarray}
E_\nu^2\Phi_{\nu_\mu}&&\approx E_\nu^2\Phi_{\bar\nu_\mu}
\approx E_\nu^2\Phi_{\nu_e}\approx
{1\over2}f_\pi I_{\rm max}\cr
&&\approx 1.5\times10^{-9}\left({f_\pi\over0.2}\right)\min\{1,E_\nu/E^b_\nu\}
{\rm GeV\,cm}^{-2}{\rm s}^{-1}{\rm sr}^{-1},\quad E^b_\nu\approx10^{14}
{\rm eV}.
\label{JGRB}
\end{eqnarray}
Here, $f_\pi$ is the fraction of energy lost to pion production by high-energy
protons. The derivation of $f_\pi$ and $E^b_\nu$ and their dependence
on GRB model parameters is given in the appendix [Eqs. (\ref{fpi}), 
(\ref{Enu})].
The intensity given by 
Eq. (\ref{JGRB}) is $\sim5$ times smaller than that given
in Eq. (8) of ref. \cite{WnB}, due to the fact that in 
Eq. (8) of ref. \cite{WnB} we neglected the logarithmic correction
$\ln(100)=5$ of Eq. (\ref{ECR}).

The GRB neutrino intensity can be estimated directly from
the observed gamma-ray fluence. The Burst and Transient Source Experiment 
(BATSE) measures the GRB fluence $F_\gamma$ over a decade of photon energy, 
$\sim0.1$MeV to $\sim1$MeV, corresponding to half a decade of radiating
electron energy (the electron 
synchrotron frequency is proportional to the square of
the electron Lorentz factor). If electrons carry a fraction $f_e$ of
the energy carried by protons, then the muon neutrino fluence of a single
burst is $E_\nu^2dN_\nu/dE_\nu\approx0.25(f_\pi/f_e)F_\gamma/\ln(3)$. 
The average neutrino flux per unit time and solid angle is obtained by
multiplying the single burst fluence with the GRB rate per solid angle,
$\approx10^3$ bursts per year over $4\pi$~sr. Using the average burst
fluence $F_\gamma\approx6\times10^{-6}{\rm erg/cm}^2$, we obtain
a muon neutrino intensity $E_\nu^2\Phi_\nu\approx3\times10^{-9}(f_\pi/f_e)
{\rm GeV/cm}^2{\rm s\,sr}$. Recent GRB afterglow observations typically
imply $f_e\sim0.1$ \cite{AG}, and therefore $f_\pi/f_e\sim1$. Thus, the 
neutrino intensity estimated directly from the gamma-ray fluence agrees with
the estimate (\ref{JGRB}) based on the cosmic-ray production rate. 

\subsection{Neutrinos at high energy $>10^{16}$~eV}

The neutrino spectrum (\ref{JGRB}) is
modified at high energy, where neutrinos are produced by the decay
of muons and pions whose life time $\tau_{\mu,\pi}$
exceeds the characteristic time for
energy loss due to adiabatic expansion and synchrotron emission 
\cite{WnB,RM98}.
The synchrotron loss time is determined by the energy density of the
magnetic field in the wind rest frame.
For the characteristic parameters of a GRB wind, the muon energy for which
the adiabatic energy loss time equals the muon life time,
$E^a_\mu$, is comparable to the energy $E^s_\mu$ 
at which the life time equals 
the synchrotron loss time, $\tau^s_\mu$. For pions, $E^a_\pi>E^s_\pi$. This,
and the fact that the adiabatic
loss time is independent of energy and the synchrotron loss time is
inversely proportional to energy, imply that
synchrotron losses are the dominant effect
suppressing the flux at high energy. The energy above which synchrotron
losses suppress the neutrino flux is
\begin{equation}
{E^s_{\nu_\mu(\bar\nu_\mu,\nu_e)}\over E^b_\nu}
\approx(\xi_B L_{\gamma,51}/\xi_e)^{-1/2}\Gamma_{300}^2
\Delta t_{\rm ms}(E^b_\gamma/1{\rm MeV})\times
\cases{10,&for $\bar\nu_\mu$, $\nu_e$;\cr 100,&for $\nu_\mu$ .\cr}
\label{sync}
\end{equation}
Here, $L_\gamma=10^{51}L_{\gamma,51}{\rm erg/s}$ is the observed gamma-ray
luminosity, $\Delta t=1\Delta t_{\rm ms}{\rm ms}$ is the observed GRB
variability time scale, $E_\gamma^b\sim1{\rm MeV}$ is the observed GRB photon
break energy, $\Gamma=300\Gamma_{300}$, and $\xi_e$ and $\xi_B$ are the 
fractions of GRB wind luminosity
carried by electrons and magnetic fields. The observational
constraints on these parameters are discussed in the appendix.
At neutrino energy $E_\nu\gg E^s_\nu$, the probability
that a pion (muon) would decay before losing its energy is approximately
given by the ratio of synchrotron cooling time to decay time
$\tau^s_{\pi(\mu)}/\tau_{\pi(\mu)}=
[E_\nu/E^s_{\nu_\mu(\bar\nu_\mu,\nu_e)}]^{-2}$,
and the intensity of Eq. (\ref{JGRB}) is suppressed by a similar factor,
\begin{equation}
E_\nu^2\Phi_{\nu_\mu(\bar\nu_\mu,\nu_e)}\approx
1.5\times10^{-9}\left({f_\pi\over0.2}\right)
\left[{E_\nu\over E^s_{\nu_\mu(\bar\nu_\mu,\nu_e)}}\right]^{-2}
{\rm GeV\,cm}^{-2}{\rm s}^{-1}{\rm sr}^{-1},\quad E_\nu\gg E_\nu^s.
\label{suppressed}
\end{equation}

Since the wind duration, i.e. the time over which energy is released from
the source, is $T\sim1$~s, 
internal shocks may occur due to variability on time scale $\delta t$
larger than the source dynamical time, 
$\Delta t\sim1{\rm ms}\le\delta t\le T\sim1$~s. 
Collisions due to variability $\delta t>\Delta t$ are less efficient in 
producing neutrinos, $f_\pi\propto\delta t^{-1}$, since the radiation energy 
density is lower at larger collision radii, leading to a smaller probability
for photo-meson interaction. However, at larger radii 
synchrotron losses cut off
the spectrum at higher energy, $E^s(\delta t)\propto\delta t$. 
Collisions at large radii therefore
result in extension of the neutrino
spectrum of Eq. (\ref{JGRB}) to higher energy, beyond the cutoff energy 
Eq. (\ref{sync}), and therefore yield
$E^2_\nu\Phi_\nu\propto E_\nu^{-1}$ for $E_\nu>E^s_\nu(\Delta t)$, since
$f_\pi\propto\delta t^{-1}\propto [E^s(\delta t)]^{-1}$. This extension is
shown in Fig. 1. We note, that on time scale $\Delta t\sim1$s the expanding
wind is expected to 
interact with the surrounding medium, driving a relativistic
shock into the ambient gas. Protons are expected to be accelerated to high
energy in this shock, 
the ``external'' shock, as well. The neutrino intensity and
spectrum produced in the external shock are given by equations 
(\ref{JGRB},\ref{fpi},\ref{sync}) with $\Delta t\sim1$s. 
Due to the low efficiency
of the external shock, $f_\pi\sim10^{-4}$, its contribution to the neutrino
flux is small. Note that as the external shock expands through a larger 
mass of the ambient gas it decelerates, and therefore on time scale 
$\Delta t\gg1$s the shock Lorentz factor is not large enough to allow
acceleration of protons to high energy.

\subsection{Comparison with other authors}

In agreement with Rachen \& M\'esz\'aros \cite{RM98}, we find that 
the neutrino flux from GRBs is small above $10^{19}$eV, 
and that a neutrino flux comparable to the $\gamma$-ray flux
is expected only below $\sim10^{17}$eV.
Our result is not in agreement, however, with that of ref. \cite{V98}, where
a much higher flux at $\sim10^{19}$eV is obtained based on the equations
of ref. \cite{WnB}, which are the same equations as used 
here.
There is a numerical error in the calculations 
of ref. \cite{V98}\footnote{\footnotesize\tightenlines The parameters 
chosen in \cite{V98} are $L_\gamma=10^{50}{\rm
erg/s}$, $\Delta t=10$s, and $\Gamma=100$. Using our equation (4) of ref.
\cite{WnB}, which is the same as Eq. (\ref{fpi}) of the present paper,
we obtain for these parameters $f_\pi=1.6\times10^{-4}$, while the author
of \cite{V98} obtains, using the same equation, $f_\pi=0.03$.}.
Finally, we note that the highest energy to which protons can be accelerated
increases with the collision radius $E_p^{\rm max}\propto\delta t^{1/3}$
\cite{GRB1}, and while $E_p^{\rm max}\sim10^{20}$eV for 
$\delta t\sim\Delta t\sim1$ms, collisions at larger radii, 
$\delta t\sim0.03$~s,
are required to allow acceleration to the highest observed energy,
$\sim3\times10^{20}$eV. In agreement with Rachen \& M\'esz\'aros \cite{RM98},
we find that at this radius the neutrino spectrum produced through
photo-meson interactions extends to $\sim10^{18}$eV [see Eqs. (\ref{sync}),
(\ref{Enu}]. 
There is no contradiction, however, between production of high-energy
protons above $\sim3\times10^{20}$eV and a break in the neutrino
spectrum at $\sim10^{16}$eV, since the efficiency of neutrino
production at collision radius corresponding to $\delta t\sim0.03$~s is
small and most of the flux is produced by collisions at smaller radii.

\section{Discussion}

We have shown that cosmic-ray observations set a model-independent upper
bound of $E^2 \Phi_\nu<2\times10^{-8}{\rm GeV/cm}^2{\rm s\,sr}$ to the 
intensity 
of high-energy neutrinos produced by photo-meson interaction
in sources 
of size not much larger than the proton photo-meson mean-free-path,
e.g. AGN jets and GRBs (see Fig. 1). 
This limit cannot be avoided  by hypothesizing
evolutionary effects of the sources (see discussion in \S IIc)
 or by invoking   magnetic field scenarios (see discussion in \S III).

Of possibly 
even greater  interest to photon astronomers, we have shown that the 
cosmic ray measurements rule out the current version of theories in which the 
gamma-ray background is due to photo-meson interactions in AGN jets.

The neutrino flux predictions of AGN jet models 
are based on two key assumptions, namely that AGN jets produce the observed
gamma-ray background and that high-energy photon emission from AGN jets is
due to decay of neutral pions produced in photo-meson interactions of
protons accelerated in the jet
to high energy. Since the neutrino flux predicted by  these
assumptions is two orders
of magnitude higher than the upper bound allowed by cosmic ray
observations
(see Fig. 1), 
at least one of the key assumptions is not valid, 
presumably the assumption that
high-energy photon emission from AGN jets is
due to photo-meson interactions.  This conclusion is 
supported by multi-wavelength observations of the
 blazar Mkn 421, which show contemporaneous strong variability at TeV and
 X-ray energies with little evidence for GeV and optical variability
  \cite{macomb95,buckley96}. This behavior suggests that
 the high-energy photon emission is due to inverse-Compton scattering by
 relativistic electrons \cite{macomb95}.  

Even if AGNs produce part of their high energy emission through
$\gamma-p$ interactions, the upper bound derived here implies
an upper bound of $\sim1$ detected AGN neutrino per year
in a high-energy neutrino telescope with an effective area of $1{\rm km}^2$
\cite{GHS}. The discussion in this paper shows that there is no 
observational motivation that would lead one  
to expect that the flux of high-energy neutrinos from AGNs will be
measurable in a ${\rm km^2}$ telescope.

The upper bound Eq. (\ref{Fmax}) also applies to the intensity of  
high-energy neutrinos that may be produced through the decay of charged pions 
created  by $p-p\rightarrow\pi^{\pm}X$ 
(rather than $p-\gamma$) interactions, as long as the 
$p-p$ optical depth in the source
is not large. At present, predictions of high-energy neutrino flux
based on such models are not available in the literature. 

The cosmic-ray flux below $3\times10^{18}$~eV is steeper than at higher
energy. This is most likely due \cite{WatsonFA} to a contribution
to the cosmic-ray flux at lower energy from Galactic sources of heavy ions. 
This view has
recently gained support from the detection in the Fly's Eye data of a
small but statistically significant enhancement of the flux of cosmic-rays
in the energy range of $2\times10^{17}$~eV to $3\times10^{18}$~eV
along the Galactic plane \cite{aniso}. 
Extra-galactic sources of cosmic-rays may therefore exist that produce
cosmic-rays with energy $<3\times10^{18}$eV at a rate higher than given by 
Eq. (\ref{ECR}). While we have no observational evidence for the
existence of such sources, we can not rule them out based on cosmic-ray
observations, and they may produce a flux of $<10^{17}$~eV neutrinos
which is higher than the upper limit implied by Eq. (\ref{Fmax}). Note that
this argument does not affect the validity of 
the upper bound (\ref{Fmax}) for AGN models, since the neutrino emission
from these sources peaks at $\sim10^{18}$~eV.

The neutrino flux predicted by the GRB model is consistent with the upper
bound derived here. The intensity estimate we give here, Eq. (\ref{JGRB}), 
is $\sim5$ times smaller
than that we gave in ref. \cite{WnB}, where the logarithmic correction of
Eq. (\ref{ECR}) was neglected. The intensity calculated here implies 
a detection rate of $\sim20$
neutrino induced muons per year for a $1\,{\rm km}^2$ detector 
(over $4\pi$~sr). As discussed in \cite{WnB}, one may look
for neutrino events in spatial and 
temporal coincidence (on a time scale of seconds)
with GRBs. 

The GRB neutrino spectrum is consistent with the secondary-particle cooling 
constraints derived by Rachen \& M\'esz\'aros \cite{RM98}. 
The neutrino flux above $\sim10^{16}$eV is suppressed, but this is also
consistent with the 
acceleration of protons to $>3\times10^{20}$eV (see Section Vc). 

Finally, we note that the 
GRB neutrino flux discussed here is the flux produced in situ, i.e. within
the source. The energy loss of high-energy protons, $>5\times10^{19}$~eV,
through photo-meson production in interaction with microwave background
photons would lead to a background neutrino intensity (which will not be
temporally associated with GRBs) comparable to the upper bound
shown in Fig. 1 at $E_\nu\sim10^{18}$~eV. This flux of high-energy neutrinos
should exist regardless of the nature of the high-energy proton
sources (assuming that these sources are indeed extragalactic, see
\cite{Yoshida}).

\paragraph*{Acknowledgments.} 

We thank the AMANDA collaboration for inviting us to attend a workshop
which stimulated this discussion. We are grateful to J. Cronin,
F. Halzen, P. Meszaros, J. P. Rachen,  
and S. Yoshida for valuable comments on an
initial version of the manuscript. This research
was partially supported by a W. M. Keck Foundation grant 
and NSF grant PHY 95-13835.

\appendix

\section{Neutrino production in GRBs}

GRBs are possible sources of high-energy cosmic-rays \cite{GRB1,GRB2}, 
which may account for the observed extra-Galactic
high-energy proton flux \cite{GRB1,CRflux}. In the GRB fireball model
\cite{fireball}, which has recently gained
support from GRB afterglow observations \cite{AG}, the observed gamma rays 
are produced by synchrotron emission of high-energy electrons accelerated in
internal shocks of an expanding relativistic wind. The hardness of
the observed spectrum, which extends to $\sim100$~MeV, requires wind 
Lorentz factors $\Gamma\sim300$ \cite{lorentz}. 
In this scenario, observed gamma-ray flux 
variability on time scale $\Delta t$ corresponds to internal collisions at 
radius $r_d\approx\Gamma^2c\Delta t$, which arise from
variability of the underlying source on the same time scale \cite{internal}.
Rapid variability time, $\sim1$~ms, observed in some GRBs \cite{ms},
and the fact that a significant fraction of bursts detected by the 
Burst and Transient Source Experiment (BATSE) show variability on
the smallest resolved time scale, $\sim10$~ms 
\cite{WL95}, imply that the
sources are compact, with linear scale $r_0\sim10^7$cm and characteristic 
dynamical time $\sim1$~ms.

In the region where electrons are accelerated, protons are also expected to be
shock accelerated, and their photo-meson interaction with observed burst
photons will produce a burst of high-energy neutrinos accompanying
the GRB \cite{WnB}. 
The neutrino spectrum is determined in this model by the observed gamma-ray
spectrum, which is well described by a broken power-law,
$dN_\gamma/dE_\gamma\propto E_\gamma^{-\beta}$ 
with different values of $\beta$ at low and high energy \cite{Band}. The
observed break energy (where $\beta$ changes) is typically 
$E_\gamma^b\sim1{\rm MeV}$, 
with $\beta\simeq1$ at energies below the break and $\beta\simeq2$ 
above the break. The interaction of protons accelerated to a power-law
distribution, $dN_p/dE_p\propto E_p^{-2}$, 
with GRB photons results in a broken power law
neutrino spectrum \cite{WnB}, $dN_\nu/dE_\nu\propto E_\nu^{-\beta}$ with
$\beta=1$ for $E_\nu<E_\nu^b$, and $\beta=2$ for $E_\nu>E_\nu^b$
(see Fig. 1). The neutrino break energy $E_\nu^b$ 
is fixed by the threshold energy
of protons for photo-production in interaction with the dominant $\sim1$~MeV
photons in the GRB,
\begin{equation}
E_\nu^b\approx5\times10^{14}\Gamma_{300}^2(E_\gamma^b/1{\rm MeV})^{-1}{\rm eV},
\label{Enu}
\end{equation}
where $\Gamma=300\Gamma_{300}$.

The normalization of the flux is determined by the
efficiency of pion production.
As shown in \cite{WnB}, the fraction of energy lost to pion production
by protons producing the neutrino flux above the break, $E^b_\nu$, is 
essentially independent of energy and is given by
\begin{equation}
f_\pi=0.20{L_{\gamma,51}\over
(E_\gamma^b/1{\rm MeV})\Gamma_{300}^4 \Delta t_{\rm ms}}.
\label{fpi}
\end{equation}
Here $\Delta t=1\Delta t_{\rm ms}{\rm ms}$ and
$L_\gamma=10^{51}L_{\gamma,51}{\rm erg/s}$ is the observed gamma-ray
luminosity. 
The values of $\Gamma$ and $\Delta t$ in Eq. (\ref{fpi})
are determined by the hardness of the $\gamma$-ray spectrum and 
by the flux variability. These parameters are also constrained by
the fact that the characteristic observed photon energy is $\sim1$~MeV.
Internal collisions are expected to be ``mildly'' relativistic in the fireball 
rest frame \cite{internal}, i.e. characterized by Lorentz factor 
$\gamma_i-1\sim1$, since adjacent shells within the wind are expected to
expand with similar Lorentz factors. 
The internal shocks would therefore heat the protons
to random velocities (in the wind frame) $\gamma_p-1\sim1$. The characteristic
frequency of synchrotron emission is determined by the characteristic energy
of the electrons and by the strength of the magnetic field. These are
determined by assuming that the fraction of energy carried
by electrons is $\xi_e$, implying a characteristic rest frame electron Lorentz
factor $\gamma_e=\xi_e(m_p/m_e)$, and that a fraction $\xi_B$ of the energy 
is carried by the magnetic field, implying 
$4\pi r_d^2c\Gamma^2B^2/8\pi=\xi_B L$ where $L$ is the 
total wind luminosity. Since the electron synchrotron cooling time is short
compared to the wind expansion time, electrons lose their energy
radiatively and $L\approx L_\gamma/\xi_e$. 
The characteristic observed energy of synchrotron photons, 
$E_\gamma^b=\Gamma\hbar\gamma_e^2 eB/m_ec$, is therefore
\begin{equation}
E_\gamma^b\approx4\xi_B^{1/2}\xi_e^{3/2}{L_{\gamma,51}^{1/2}
\over\Gamma_{300}^2\Delta t_{\rm ms}}{\rm MeV}. 
\label{Eg}
\end{equation}
At present, there is no theory that allows the determination of 
the values of the equipartition fractions $\xi_e$ and $\xi_B$. However,
for values close to equipartition, the model photon break energy
$E_\gamma^b$ is consistent with the observed $E_\gamma^b$ for $\Gamma=300$
and $\Delta t=1$~ms.


\begin{references}
\tightenlines
\bibitem{GHS} T. K. Gaisser, F. Halzen, and T. Stanev, Phys. Rep.
 {\bf 258}, 173 (1995).
\bibitem{WatsonFA} A. A. Watson, Nucl. Phys. B (Proc. Suppl.) {\bf 22B}, 116 
 (1991); D. J. Bird {\it et al.}, Phys. Rev. Lett. {\bf 71}, 3401 (1993);
 S. Yoshida {\it et al.}, Astropar. Phys. {\bf 3}, 151 (1995).
\bibitem{WnB} E. Waxman and J. N. Bahcall, 
 Phys. Rev. Lett. {\bf 78}, 2292 (1997).
\bibitem{AGNnu} K. Mannheim, Astropar. Phys. {\bf 3}, 295 (1995); 
  F. Halzen and E. Zas, Astrophys. J. {\bf 488}, 669 (1997);
  R. J. Protheroe, Adelaide preprint ADP-AT-96-4 (astro-ph/9607165).
\bibitem{RM98} J. P. Rachen and P. M\'esz\'aros, Submitted to Phys. Rev.
 D (1998) (astro-ph/9802280).
\bibitem{Stecker} F. Stecker, C. Done, M. Salamon, and P. Sommers,
 Phys. Rev. Lett. {\bf 66}, 2697 (1991); erratum 
 Phys. Rev. Lett. {\bf 69}, 2738 (1992).
\bibitem{CRflux} E. Waxman, Astrophys. J. {\bf  452}, L1 (1995).
\bibitem{aniso} D. J. Bird {\it et al.}, Submitted to the Astrophys. J.
 (1998) (astro-ph/9806096).
\bibitem{QSO} B. J. Boyle and R. J. Terlevich, Mon. Not. 
 Roy. Astron. Soc.  {\bf 293}, L49 (1998).
\bibitem{QSOl} P. C. Hewett, C. B. Foltz and F. Chaffee, Astrophys. J. 
 {\bf 406}, 43 (1993).
\bibitem{QSOh} M. Schmidt, D. P. Schneider, and J. E. Gunn, Astron. J.
 {\bf 110}, 68 (1995).
\bibitem{SFR} S. J. Lilly, O. Le Fevre, F. Hammer and D. Crampton,
 Astrophys. J.  {\bf 460}, L1 (1996); P. Madau, H. C. Ferguson,
 M. E. Dickinson, M. Giavalisco, C. C. Steidel and A. Fruchter,
 Mon. Not. Roy. Astron. Soc.  {\bf 283}, 1388 (1996).
\bibitem{field1} 
  P. P. Kronberg, Rep. Prog. Phys.  {\bf 57}, 325 (1994).
\bibitem{field2}
  J. P. Vallee, Astrophys. J. {\bf 360}, 1 (1990).
\bibitem{Kim91}
  K. T. Kim, P. C. Tribble, and P. P. Kronberg, Astrophys. J. 
  {\bf 379}, 80 (1991).
\bibitem{Kulsrud}  R. M. Kulsrud, R. Cen, J. P. Ostriker, and D. Ryu,
  Astrophys. J. {\bf 480}, 481 (1997).
\bibitem{Ryu} D. Ryu, H. Kang, and P. L. Biermann, Astron. Astrophys.
  {\bf 335}, 19 (1998).
\bibitem{Biermann} G. Sigl, M. Lemoine, and P. Biermann, Astropar. Phys.
  in press (astro-ph/9806283).
\bibitem{Tanco} G. A. Medina-Tanco, Astrophys. J. in press (astro-ph/9808073).
\bibitem{Strauss} M. A. Strauss, and J. A. Willick, Phys. Rep. {\bf 261},
  271 (1995).
\bibitem{pionAGN} P. L. Biermann, \& P. A. Strittmatter, Astrophys. J.
 {\bf 322}, 643 (1987); K. Mannheim, Astron. Astrophys. {\bf 269}, 67
 (1993).
\bibitem{TeV} J. E. McEnery {\it et al.}, 25th Int. Cosmic Ray Conference, 
 Durban 1997 (astro-ph/9706125); R. J. Protheroe {\it et al.} 25th Int. 
 Cosmic Ray Conference, Durban 1997 (astro-ph/9710118).
\bibitem{EGRET} D. J. Thompson, \& C. E. Fichtel, Astron. Astrophys. 
 {\bf 109}, 352 (1982).
\bibitem{eTeV} M. Punch {\it et al.} Nature {\bf 358}, 477 (1992);
 J. Quinn {\it et al.} Astrophys. J. {\bf 456}, L83 (1996); 
 S. M. Bradbury {\it et al.}, Astron. Astrophys. {\bf 320}, L5 (1997).
\bibitem{fireball} For a recent review see 
  T. Piran, in Unsolved Problems In Astrophysics, eds.
  J. N. Bahcall and J. P. Ostriker (Princeton: Princeton Univ. Press), 343-377 
  (1996).
\bibitem{AG} E. Waxman,  Astrophys. J. {\bf 485}, L5 (1997); 
 A. M. J. Wijers, M. J. Rees, and P. M\'esz\'aros, Mon. Not. 
 Roy. Astron. Soc.  {\bf 288}, L51 (1997); E. Waxman, S. Kulkarni, and
 D. Frail, Astrophys. J. {\bf 497}, 288 (1998).
\bibitem{lorentz} J. H. Krolik and E. A. Pier, Astrophys. J. 
 {\bf  373}, 277 (1991); M. G. Baring and A. K. Harding, Astrophys. J. 
 {\bf 491}, 663 (1997).
\bibitem{internal} M. Rees, and  P. M\'esz\'aros, Astrophys. J. {\bf 430}, 
 708 (1994);
 B. Paczy\'nski, and G. Xu, Astrophys. J. {\bf 427}, 708 (1994).
\bibitem{GRB1} E. Waxman, Phys. Rev. Lett. {\bf 75}, 386 (1995).
\bibitem{GRB2}
 M. Milgrom, and V. Usov, Astrophy. J. {\bf 449}, L37 (1995); M. Vietri, 
 Astrophys. J. {\bf 453}, 883 (1995).
\bibitem{V98} M. Vietri, Phys. Rev. Lett. {\bf 80}, 3690 (1998).
\bibitem{macomb95}D. J. Macomb {\it et al.}, Astrophys. J. {\bf 449},
L99 (1995); erratum Astrophys. J. {\bf 459}, L111 (1996). 
\bibitem{buckley96}J. H. Buckley {\it et al.}, Astrophys. J. {\bf 472}, L9 
(1996).
\bibitem{Yoshida} S. Yoshida and M. Teshima, Prog. Theor. Phys. {\bf 89},
 833 (1993).
\bibitem{ms} P. N. Bhat {\it et al.}, Nature {\bf 359}, 217 (1992).
\bibitem{WL95} E. Woods and A. Loeb, Astrophys. J. {\bf  453}, 583 (1995).
\bibitem{Band} D. Band {\it et al.}, Astrophys. J. {\bf  413}, 281 (1993).
\end{references}
\end{document}